\documentclass[aps,prb,preprint,groupedaddress,showpacs]{revtex4}
\usepackage{amsmath,amssymb,amsfonts}
\usepackage{graphicx}
\DeclareGraphicsExtensions{.jpg,.pdf,.png,.eps}
\newcommand{\mW} {$m\Omega$ cm }
\newcommand{\SFA} {SrFe$_{2}$As$_{2}$ }
\newcommand{\BFA} {BaFe$_{2}$As$_{2}$ }
\newcommand{\AFA} {AEFe$_{2}$As$_{2}$ }
\newcommand{\SFAf} {SrFe$_{2}$As$_{2}$}
\newcommand{\BFAf} {BaFe$_{2}$As$_{2}$}

\newcommand{\CFA} {CaFe$_{2}$As$_{2}$ }
\newcommand{\CFAf} {CaFe$_{2}$As$_{2}$}

\begin{document}
\title{Complete pressure dependent phase diagrams for \SFA and \BFA}

\author{E. Colombier}
\author{S. L. Bud'ko}
\author{N. Ni}
\author{P. C. Canfield}
\affiliation{Ames Laboratory and Department of Physics and Astronomy, Iowa State University, Ames, IA 50011, USA}

\date{\today}

\begin{abstract}
The temperature dependent electrical resistivity of single crystalline \SFA and \BFA has been measured in a liquid medium, modified Bridgman anvil cell for pressures in excess of 75~kbar. These data allow for the determination of the pressure dependence of the higher temperature, structural / antiferromagnetic phase transitions as well as the lower temperature superconducting phase transition. For both compounds the ambient pressure, higher temperature structural / antiferromagnetic phase transition can be fully suppressed with a dome-like region of zero resistivity found to be centered about its critical pressure. Indeed, qualitatively, the temperature dependence of the resistivity curves closest to the critical pressures are the closest to linear, consistent with possible quantum criticality.  For pressures significantly higher than the critical pressure the zero resistivity state is suppressed and the low temperature resistivity curves asymptotically approach a universal, low temperature manifold. These results are consistent with the hypothesis that correlations / fluctuations associated with the ambient-pressure, high-temperature, tetragonal phase have to be brought to low enough temperature to allow superconductivity, but if too fully suppressed can lead to the loss of the superconducting state.
\end{abstract}


\maketitle 
\section{Introduction}
Pressure is a thermodynamic variable that can be used to perturb and, hopefully, understand phase transitions. In the case of the recently discovered families of FeAs-based superconductors pressure has been particularly powerful. The application of pressure on polycrystalline LaFeAs(O/F) raised the onset of the superconducting transition from 26~K to 43~K \cite{Takahashi08} and the substitution of heavier rare earths for La achieved even larger ambient pressure increases in T$_{c}$ to 55 K via the lanthanide contraction (chemical pressure) \cite{Ren08}.
Even pure LaFeAsO has been found to superconduct with a maximum T$_{c}$=21~K for a 120~kbar pressure \cite{Okada08}. After the discovery of superconductivity in K-doped \AFA (AE = Ba and Sr) \cite{Rotter08,Chen08CPL}, \CFA was discovered as a new member of the \AFA series and found to have the smallest lattice parameters \cite{Ni08}. The application of pressure to \CFA using self clamping cells with liquid media lead to the discovery of superconductivity for P $\sim$ 5 kbar \cite{Torikachvili08Ca,Park08}. Subsequent work with He-medium cells \cite{Yu09,Goldman09,Kreyssig08} demonstrated that (i) these materials are not only exceptionally pressure sensitive, but can also be very sensitive to strain and (ii) in the case of \CFAf , where there is a pressure stabilized collapsed tetragonal phase at low temperatures for P $\geqslant$ 5 kbar, the combination of a solidified pressure medium and a first order, structural phase transition leads to a multi-crystallographic phase at low temperatures with the superconductivity most likely coming from a remaining orthorhombic or even high temperature tetragonal phases. Subsequent studies of the effects of pressure on \SFA and \BFA have found that the higher temperature, structural / antiferromagnetic phase transition is much less pressure sensitive and that much higher pressures are needed to stabilize superconductivity, but there is poor agreement between the differing measurements, using differing pressure environments, often measuring only the upper, structural / antiferromagnetic or the lower, superconducting, phase line \cite{Torikachvili08,Alireza08,Igawa08,Kotegawa09,Fukazawa08,Mani09condmat}. In addition, strain stabilized superconductivity seems to be an ubiquitous feature of all of the \AFA materials \cite{Saha08condmat,Torikachvili08Ca}.

In this paper we assemble and present pressure - temperature phase diagrams for \BFA and \SFA using a liquid medium modified Bridgman cell configuration to measure temperature and pressure dependent electrical resistivity of multiple samples of each compound. We are able to determine both the pressure dependence of the upper, structural / antiferromagnetic phase transition, as well as the lower, superconducting phase. We have found that whereas there is a broad region of partial superconductivity (not even completely filamentary) that gives rise to a partial electrical shorting of the sample, zero resistivity exists over a smaller pressure range that is centered on the extrapolated termination of the higher temperature structural / antiferromagnetic phase transition line.

\section{Experimental details}
The \SFA and \BFA samples studied here were single crystals grown out of tin \cite{Yan08}, and FeAs flux \cite{Ni08Codoped} respectively. Electrical resistivity was measured by a four-probe method. Samples were cleaved and then cut to appropriate dimensions (typically 700 $\times$ 150 $\times$ 30~$\mu m^{3}$) for high pressure studies. Four 12.5~$\mu m$ diameter gold wires were fixed with silver epoxy in the (\textit{a},\textit{b}) plane.

\begin{figure}
\begin{center}
\includegraphics[angle=0,width=120mm]{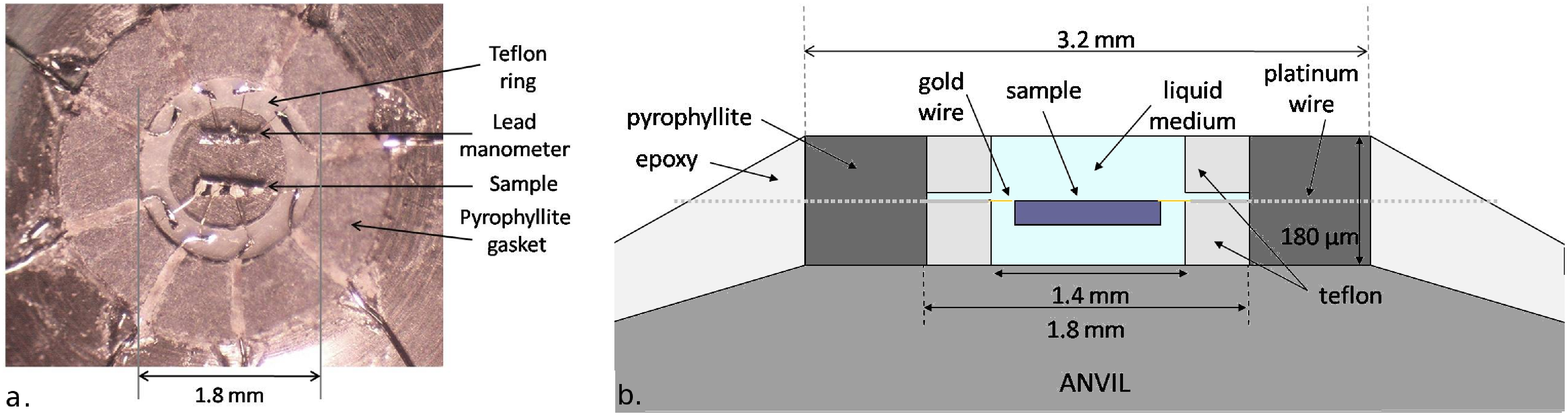}
\end{center}
\caption{(Color online) (a) Top view (photograph) of the pressure chamber with a spot-welded lead manometer (top) and a \SFA sample (bottom) to which wires were fixed with silver epoxy. A second teflon ring is then placed on top of the first one, just before filling with liquid. (b) Sketch of the side view.}
\label{cellB2}
\end{figure}

We performed resistivity measurements under pressures of up to 76 kbar with a Bridgman cell modified to use a liquid pressure medium \cite{Colombier07}; a Fluorinert mixture 1:1 FC70:FC77 was chosen in the present case. The pressure was determined at low temperature by the superconducting temperature transition of a lead sample \cite{Bireckoven88}. The top and side views of such a pressure chamber are represented in figure \ref{cellB2} both as a photograph and as a schematic.

All temperature and field dependent measurements were performed in a Quantum Design Physical Property Measurement System (PPMS). A standard PPMS sample puck was fixed to a small pressure cell (23~mm diameter, 60~mm length, and 130~g mass). No thermometers were placed on the pressure cell. We chose a measurement current of 1~mA, although we measured the superconducting transitions for different currents: 1~mA, 0.1~mA and sometimes 0.01~mA, as needed. The magnetic field dependence of the superconducting transition was measured up to 14 T, with field along the \textit{c} axis of the samples. The typical pressure variations between ambient and low temperature were previously estimated to be lower than 1~kbar, by fitting our lead data with a Bloch-Gr\"uneisen law as proposed Eiling and Schilling \cite{Eiling81}.

\begin{figure}
\begin{center}
\includegraphics[angle=0,width=80mm]{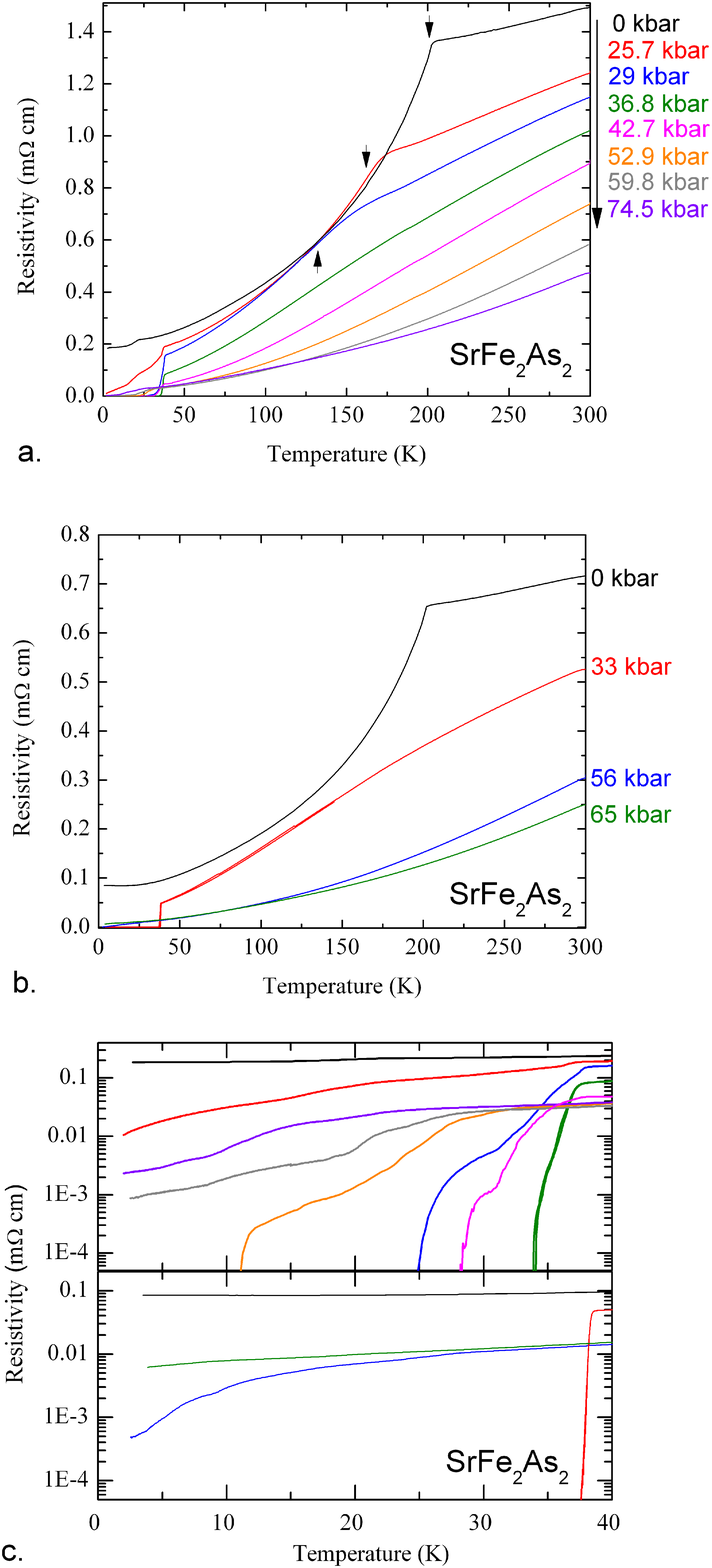}
\end{center}
\caption{(Color online) (a) and (b) summarize the resistivity measurements under pressure for two different \SFA samples: cell 1 and cell 2 respectively. (c) presents an enlargement of the low temperature behavior with resistivity on logarithmic scale (from $5\times 10^{-5}$ to 0.4~\mW and 0.2~\mW respectively for cell 1 and cell 2). Arrows show the structural / antiferromagnetic transition temperature deduced from a maximum of the resistivity derivative criterion.}
\label{cell}
\end{figure}

\begin{figure}
\begin{center}
\includegraphics[angle=0,width=80mm]{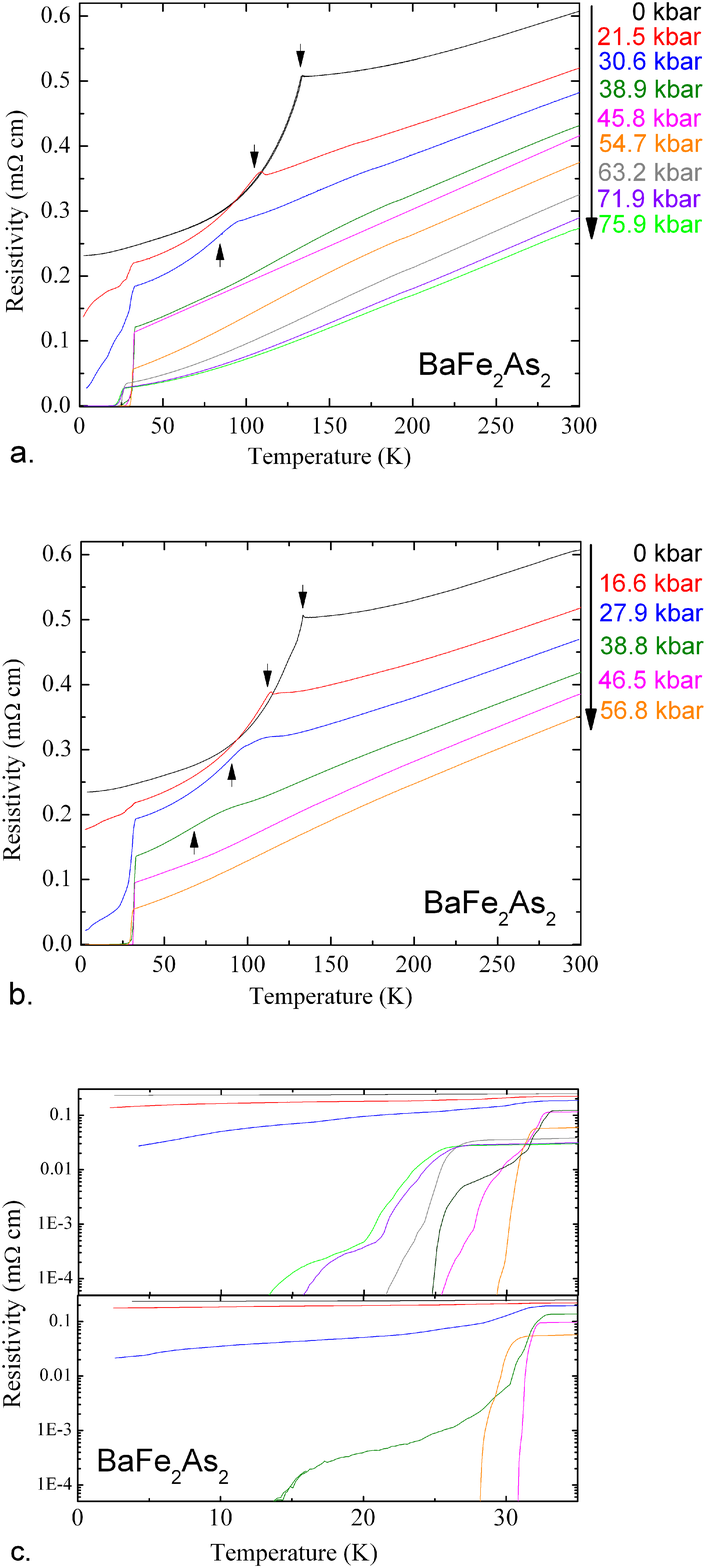}
\end{center}
\caption{(Color online) (a) and (b) summarize the resistivity measurements under pressure for two different \BFA samples. Two sets of measurements are respectively called cell 1 and cell 2. For cell 1, pressure uncertainty is around 3~kbar for the two last pressures, due to a small remanent field. The arrows show the structural / antiferromagnetic transition temperature deduced from our criterion (maximum of the resistivity derivative). (c) presents an enlargement of the low temperature behavior with resistivity in logarithmic scale from $5 \times 10^{-5}$ to 0.3~\mW . The higher and lower panels correspond to cell 1 and cell 2, respectively.}
\label{cellBa}
\end{figure}

Measurements between 40 and 300~K were performed at constant cooling or heating rates. The comparison between subsequent measurements of the resistivity of a lead sample measured first inside and then outside of the pressure cell (filled with 1:1 FC70:FC77) at ambient pressure gave us an estimate of the temperature shift between the PPMS sample thermometer and the temperature of our pressurized samples. The comparison under pressure for the resistivity difference between warming and cooling gave similar results. A cooling rate of 1~K / min results in a nearly uniform temperature shift lower than 3~K between around 290 and 80~K. We used a slower temperature sweep of 0.5~K / min during warming for more precise measurements. In this case, the shift between the real pressure cell temperature and the measured one is still nearly uniform and was estimated to be lower than 1.2~K. Most of the data shown in figures \ref{cell} and \ref{cellBa} were obtained on warming at a rate of 0.5~K / min. For temperatures below 40~K, a 0.2 K / min rate results in a temperature shift lower than 100~mK. To precisely measure the low temperature, superconducting transitions of the lead and the sample, the temperature was stabilized before the measurement of each data point. No shift in the data between cooling and warming was observed by proceeding this way.

\begin{figure}
\begin{center}
\includegraphics[angle=0,width=120mm]{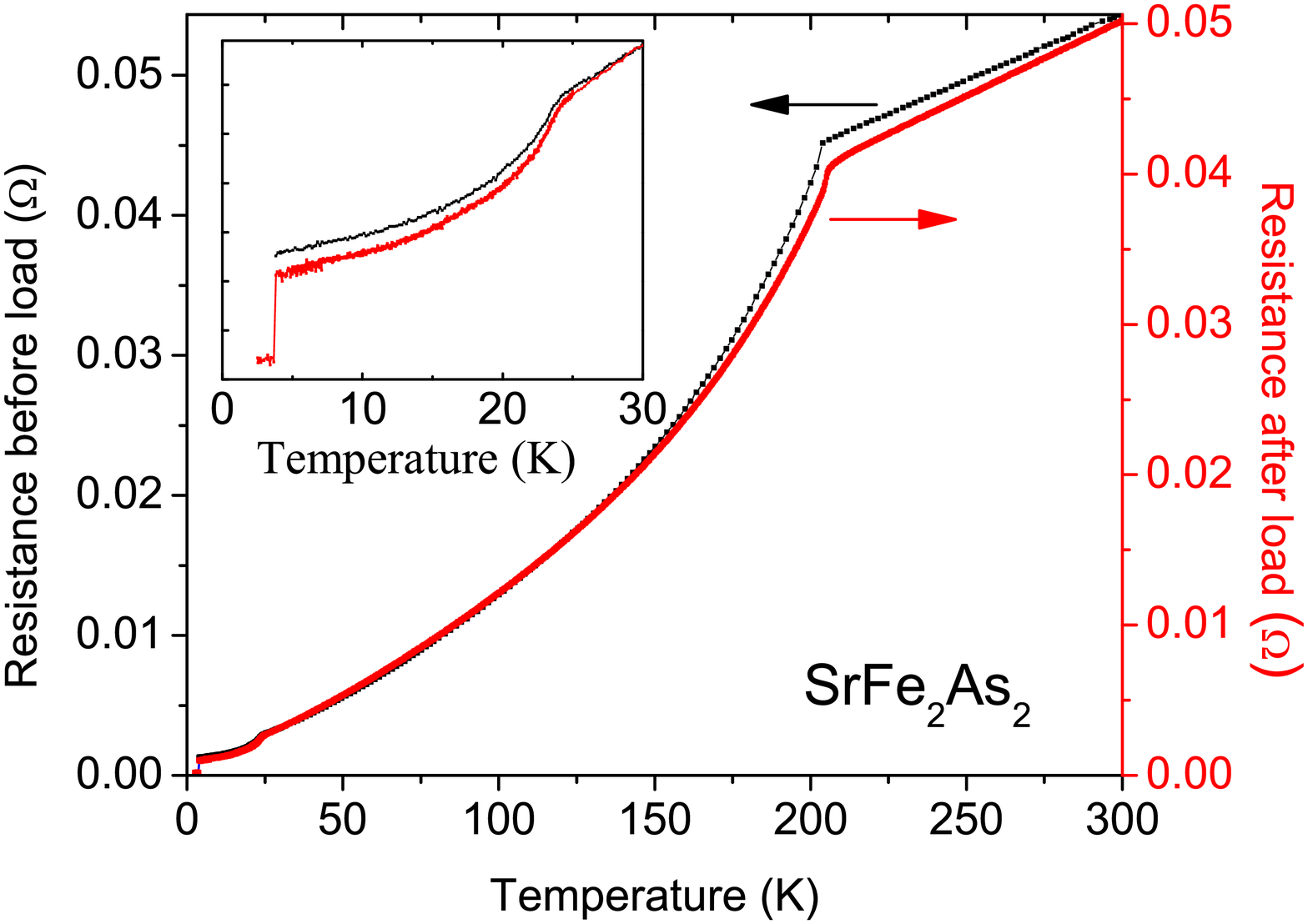}
\end{center}
\caption{(Color online) Resistivity at ambient pressure for a \SFA sample before and after a pressure cycle up to around 50~kbar. Two different resistivity scales were used for a better comparison. A low temperature zoom is shown in the inset.}
\label{beforafterload}
\end{figure}

Three pressure cells with \SFA samples and two with \BFA samples were measured so as to check the reproducibility of their behavior under pressure. For these pressure cells, measured down to low temperatures up to high pressures, we were not able to gently remove the sample after the final measurement. (The gasket would break on decompression, resulting in the loss of the sample. Some details about ruptures during unloads are given by Colombier and Braithwaite \cite{Colombier07}.) To show that the pressure conditions were not harmful for the samples, we applied to a \SFA sample a pressure estimated to be around 50~kbar. This pressure cell was not measured at low temperature, but was kept at 300~K for one night before careful and successful unloading. The comparison at ambient pressure, before and after the load on this \SFA sample is shown in figure \ref{beforafterload}. The general behavior remains the same and the resistance didn't increase after the pressure unload compared to before, indicating that no cracks or irreversible defects appeared. Some silver paste was added to the sample contacts, weakened during the unload, which might explain the slight differences between the curves. The distance between voltage wires became indeed around 10~$\%$~smaller, which could have increased the relative contribution of tin, and might have also caused the difference between the two curves, around the structural transition temperature.

\begin{figure}
\begin{center}
\includegraphics[angle=0,width=120mm]{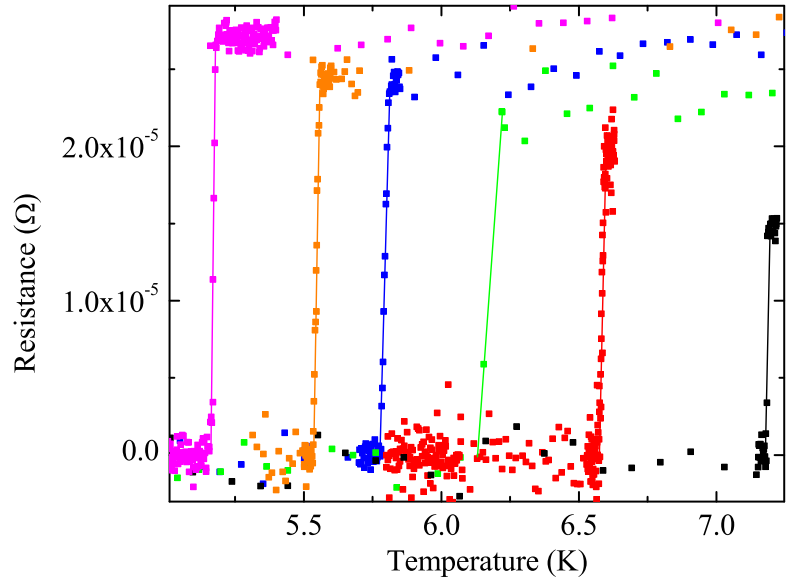}
\end{center}
\caption{(Color online) Lead resistance at low temperature from one pressure cell. The superconducting transition is shown at six different pressures. From right to left: 0 kbar, 16.6 kbar, 27.9 kbar, 38.8 kbar, 46.5 kbar and 56.8 kbar}
\label{LeadTransition}
\end{figure}

We may however worry that the measured samples could be damaged during cooling, because of thermal contractions of the pressure cell, even if the pressure variations estimated are low. Moreover, when the structural transition occurs, the sample dimensions may suddenly change by a few percent (in particular along the \textit{a} and \textit{b} axes \cite{Yan08,Rotter08PRB}) and it will be strained by a solidified (frozen) medium. This was underlined in the case of \CFA from studies using pressure mediums presenting different hydrostatic conditions \cite{Torikachvili08Ca,Park08,Yu09,Lee08condmat}. As can be seen in figure \ref{LeadTransition}, good hydrostatic conditions can be inferred from the narrow superconducting transition of the lead, even though the pressure medium has frozen by this temperature. The pressure gradients between voltage wires, estimated from the superconducting width (between the true onset and the zero-resistivity temperatures), are less than 0.6 kbar at 56.8 kbar. (This corresponds to a 20~mK superconducting transition width, which is relatively small.) For the higher pressures, we obtain typical widths from 20 to 40~mK. Some typical values of pressure gradient estimations and other tests regarding the pressure quality are given by Colombier and Braithwaite \cite{Colombier07}.

Samples were characterized at ambient pressure by resistivity measurement between 2 and 300~K. The typical residual resistivity ratio was $RRR_{~2-300\text{K}}\equiv\dfrac{\rho_{~300\text{K}}}{\rho_{~2\text{K}}}\thickapprox 8$ for \SFA and around 2.6 for \BFAf . Some differences were observed between the \SFA samples, specifically for the low temperature behavior. In particular, we noticed in many samples a kink around 21.5~K (clearly shown in the inset of figure \ref{beforafterload}), with an amplitude ranging from a few percent to 75~$\%$ of the resistivity value. This anomaly has been observed at ambient pressure by several groups \cite{Saha08condmat,Torikachvili08}. It is attributed to small regions of superconductivity and may be created by internal strains \cite{Saha08condmat}. Some samples also presented a partial superconducting transition, at around 3.7~K, due to the presence of tin flux (again shown in the inset of figure \ref{beforafterload}). We avoided such tin-contaminated samples for high pressure measurements.

We observe the changes under pressure in the resistive signature of the transition attributed to the combined structural and antiferromagnetic transition, up to around 30 and 40~kbar respectively for \SFA and \BFAf . Whereas both compounds manifest a similarly shaped resistive signature at ambient pressure, a small, sharp, but clearly detectable, local maxima is seen just above the loss of resistivity under pressure for \BFA samples but not for \SFA (as can be seen in figures \ref{cell} and \ref{cellBa} below). This feature may be attributed to a superzone gap opening and the fact that it is only observed in some of the samples may be due to in-plane anisotropy \cite{Sergey00}. However, this feature seems to be linked to the samples batches much more than to the compound. Although we didn't observe any such feature in the three \SFA samples measured, Kotegawa \textit{et al.} \cite{Kotegawa09} saw this feature under pressure for their Sn-grown single crystals of \SFAf . By carefully examining resistivity curves from our samples and from other \SFA studies \cite{Kotegawa09,Torikachvili08,Kumar08}, it seems that this feature develops under pressure only for samples presenting already a sharp peak of small amplitude (less than 1~\%) at ambient pressure. In our \BFA samples, such a peak was also observed at ambient pressure for most of the samples measured.

\section{Results}

\subsection{\SFA}
Figure \ref{cell} presents temperature dependent resistivity under pressure for two different \SFA samples. The difference between the 300~K resistivity values at ambient pressure is most likely due to a combination of uncertainty in determination of sample dimensions (up to ten percent in the basal plane and up to 30 percent for the thickness) and cracks or defects induced while cleaving the samples, which might change the current path.

\begin{figure}
\begin{center}
\includegraphics[angle=0,width=120mm]{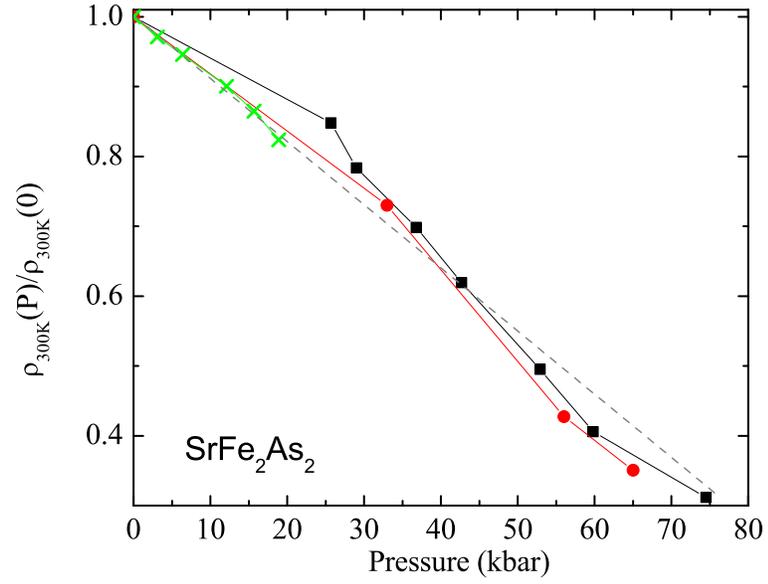}
\end{center}
\caption{(Color online) Relative resistivity decrease at 300~K  $\rho_{300\text{K}}(P)/\rho_{300\text{K}}(0)$ versus pressure from cell 1 (black squares) and cell 2 (red circles). Green crosses up to 20~kbar are data from reference [\onlinecite{Torikachvili08}]. The dashed line is a guide for the eye.}
\label{R300KSr}
\end{figure}

The relative resistivity decrease at ambient temperature is shown in figure \ref{R300KSr} for our two \SFA samples. For comparison, low pressure data using a piston-cylinder cell \cite{Torikachvili08} were added. The three sets of measurements are in a quite good agreement. If any damage occurs to the samples during the temperature cycle it results in small resistivity differences compared to the pressure induced changes. We observed a resistivity decrease in a monotonic and essentially linear fashion with pressure, with a close to a factor 2.5 decrease between 0 and 65~kbar.

The drop in resistance that is associated with the combined structural and antiferromagnetic transitions, is observed around 202~K at ambient pressure. The transition, relatively sharp at ambient pressure, is shifted to lower temperatures and becomes broader and less pronounced as the pressure is increased. We could not observe it clearly for pressures higher than 30~kbar.

When we apply pressure, a kink appears at 37.6~K. As pressure is increased, it becomes more pronounced and for pressures higher than 29~kbar the kink becomes a complete transition to zero resistivity. This transition progressively becomes narrower up to around 35 kbar. For higher applied pressures it is then broadens again and shifted down to lower temperatures.

The low pressure kink at low temperature has been observed previously by Torikachvili \textit{et al.} \cite{Torikachvili08} but as described above, it did not become a clear transition to a zero-resistivity state by the maximum applied pressure of 19 kbar. Its transition temperature is in very good agreement with the one we observed at ambient pressure for the sample measured in cell 1. Up to 16~kbar, the onset temperature of this feature remains relatively constant and at 18.9~kbar, there is an increase in the onset temperature of this feature.

\begin{figure}
\begin{center}
\includegraphics[angle=0,width=120mm]{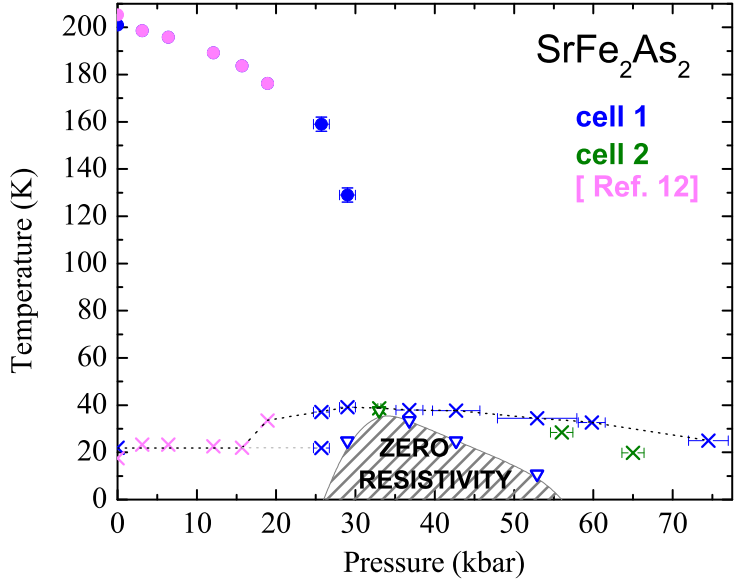}
\end{center}
\caption{(Color online) Phase diagram T(P) of \SFA deduced from resistivity measurements in a modified Bridgman pressure cell. Blue and green data correspond to cell 1 and cell 2, respectively. Low pressure data up to 20~kbar from reference [\onlinecite{Torikachvili08}] were added in pink. Circles and crosses correspond respectively to the structural / antiferromagnetic transition and the the onset of superconductivity. Vertical error bars indicates the shift between cooling and warming, due to a not perfect thermalization. Triangles represent the offset temperature of the full superconducting transition. The hatched area shows the zero-resistance superconducting region. The very large horizontal error bars are due to pressure uncertainties caused by a small remanent field. These uncertainties were estimated from one pressure (60~kbar), where the pressure cell was measured in two different PPMS, one without remanent field. Smaller error bars were estimated from the superconducting width, indicating at the same time pressure gradients and a not perfect thermalization. Dotted lines link the onset temperature of the superconducting transition.}
\label{DP}
\end{figure}

Figure \ref{DP} presents the phase diagram which summarizes our measurements (up to above 70~kbar) together with data from Torikachvili \textit{et al.} \cite{Torikachvili08} study up to 19~kbar. We defined the antiferromagnetic / structural transition temperature as the maximum of the resistivity derivative, $\dfrac{d\rho}{dT}$. The onset temperature of the down-turn in resistivity (the kink) was chosen as T$_{c}$ . The temperature below which zero resistance was measured is also shown. As the transition sharpens for P $\sim$ 35~kbar, these two temperatures approach each other.

\begin{figure}
\begin{center}
\includegraphics[angle=0,width=120mm]{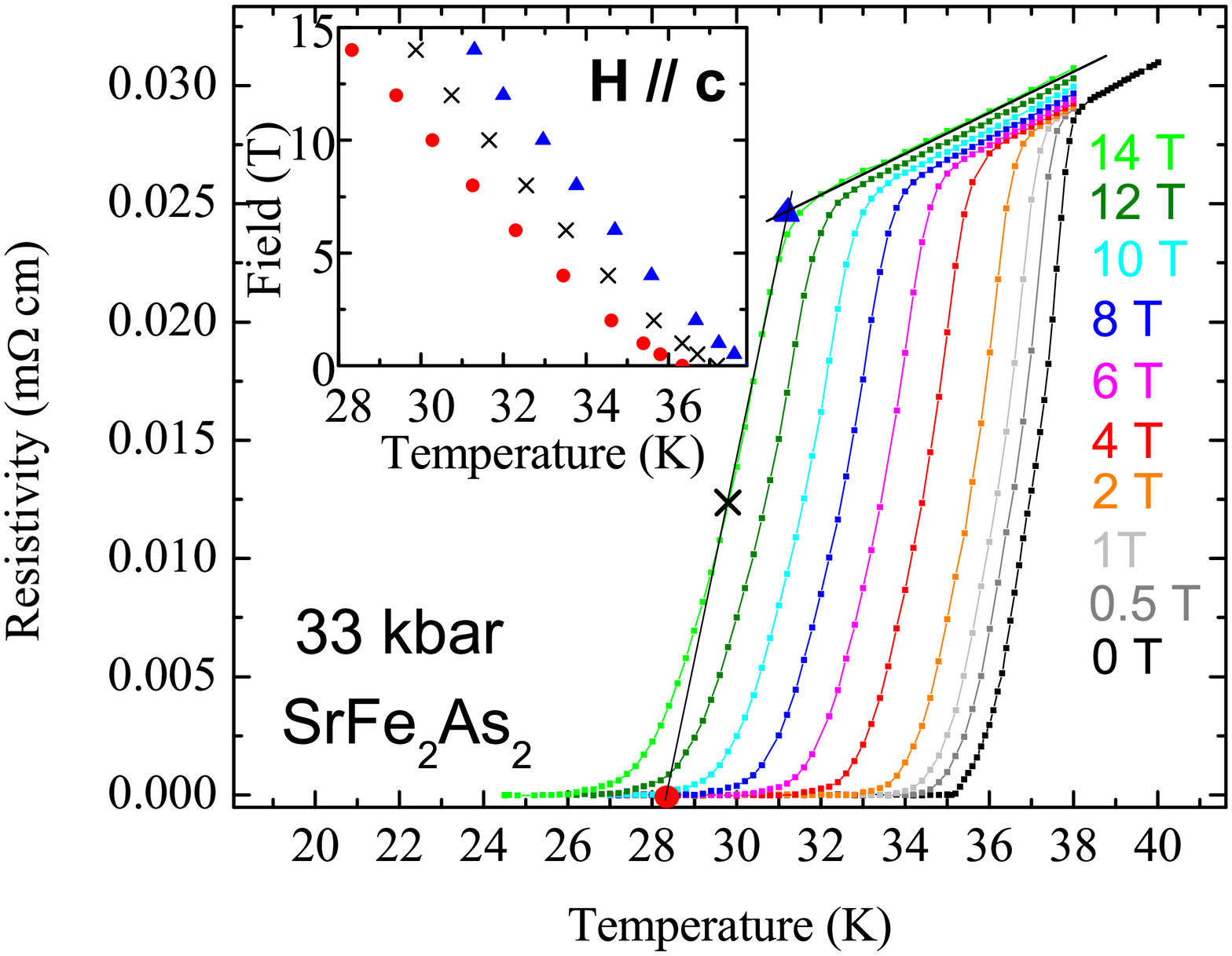}
\end{center}
\caption{(Color online) Resistivity of \SFA under different magnetic fields measured up to 14~T. The criterions used to determine the onset, half width and offset temperatures are shown for the superconducting transition at 14~T. The insert summarizes the field dependence of the onset (triangles), the offset (circles) and the half width (crosses) temperatures.}
\label{field}
\end{figure}

Figure \ref{field} presents the effects of an applied magnetic field along the \textit{c}-axis on the superconducting transition. Kotegawa \textit{et al.} \cite{Kotegawa09} measured the resistivity in field at 41.5~kbar and found a decrease for T$_{c}$ from 30~K at 0~T to 27~K at 8~T , and H$_{c2}$(0~K) around 86~T (from a linear extrapolation). For P $\sim$ 33~kbar, we found 60~T $\leqslant$ H$_{c2}$(0~K) $\leqslant$ 80~T depending on the criterion for H$_{c2}$ and the extrapolation used. It should be noted that the transition width (up to 14~T) is not very sensitive to the magnetic field. The transition width increases from 1.7~K in H = 0~T to 3~K in H = 14~T.

\subsection{\BFA}
Figure \ref{cellBa} presents the general behavior of \BFA samples under pressure. The resistivity data from two different samples are shown for comparison.
For both sets of measurements, we observe an evolution of the resistive signature of the ambient pressure 130~K structural / antiferromagnetic phase transition with pressure. For low pressure there is a sharp loss of resistance, preceded by a small (superzone gap-like) local maximum \cite{Sergey00}. As pressure increases, the loss of resistance decreases and the local maximum broadens and weakly increases. By 40~kbar, both features are no longer detectable. In order to be consistent with the \SFA study, we chose the maximum of resistivity derivative as a criterion to determine the transition temperature.

Unlike \SFAf , at ambient pressure there is no low temperature kink, but as soon as pressure is applied, one appears, with an onset temperature around 33~K. As pressure is increased the drop in resistance sharpens although the onset temperature of this kink-like feature does not change significantly. By 40~kbar a zero resistance state is stabilized, reaching a maximum for applied pressure near 55~kbar. At higher applied pressures the onset of the kink decreases slightly and the width of the transition broadens again.

The general behavior observed for both pressure cells (Fig. \ref{cellBa} as well as Figs. \ref{R300KBa} and \ref{DPBa} below) shows good reproducibility for \BFA under high pressure.
Figure \ref{R300KBa} shows that the resistivity at ambient temperature for cells 1 and 2 present a very similar decrease. As was the case for \SFA (Fig. \ref{R300KSr}) the decrease is essentially linear over this pressure range.

It should be noted that for temperatures higher than 100~K, we observe a few differences between samples, such as the resistivity slope, steeper in the case of cell 1 for all pressures. The  anisotropy reported for this compound is considered as small, but we might expect from anisotropic studies \cite{Tanatar08} that it is enough to show changes in resistivity when the current path is different.

Figure \ref{DPBa} presents the phase diagram obtained for the \BFA samples. The temperature we inferred for the combined structural and antiferromagnetic transitions decreases almost linearly (but with a slight curvature), with a -2.2~K / kbar slope. The zero-resistivity phase appears in a pressure range where the high temperature transition is still observed. There is very little variation in the temperature of the kink onset. On the other hand the zero-resistivity region is much more pressure dependent. For pressures near 50~kbar these temperatures become close and there is a relatively sharp, single transition to the superconducting state.

\begin{figure}
\begin{center}
\includegraphics[angle=0,width=120mm]{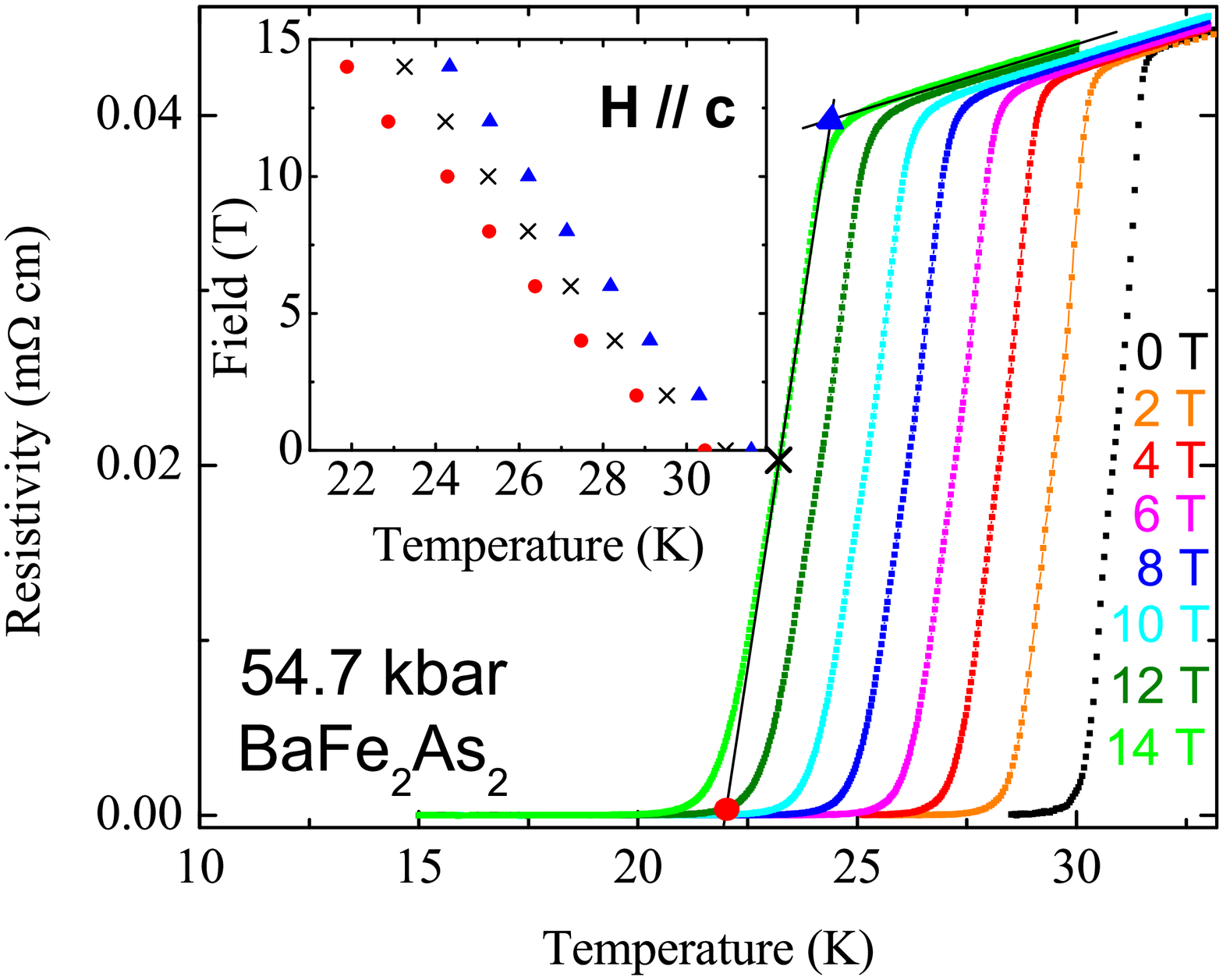}
\end{center}
\caption{(Color online) Resistivity of \BFA under different magnetic fields measured up to 14~T. The criterions used to determine the onset, half width and offset temperatures are shown for the superconducting transition at 14~T. The insert summarizes the field dependence of the onset (triangles), the offset (circles) and the half width (crosses) temperatures.}
\label{Bafield}
\end{figure}

The superconducting transition of \BFA in magnetic field up to 14~T is shown figure \ref{Bafield}. The pressure measured is 54.7~kbar, in the region where the superconducting transition is narrow. The behavior observed is very similar to \SFAf, whereas the H$_{c2}$ we estimated from this relatively low field range is around 20~T lower. The transition width increases from 1.1~K in H = 0~T to 2.5~K in H = 14~T. In particular, the zero-resistance state is obtained more slowly. An increase of the T$_{c,onset}$ slope is also noticed. From a linear extrapolation to T~=~0~K of the three data points with H$\geq$10~T, we obtain H$_{c2}$=66~T whereas H$_{c2}$=60~T if we use data with 4~T$\leq$H$\leq$8~T.

\section{Discussion}
This comprehensive study of the response of both the structural / antiferromagnetic and superconducting transitions of single crystalline \SFA and \BFA to hydrostatic pressure can be compared to earlier, partial studies of these compounds. In the case of \SFA lower pressure transport measurements by Torikachvili \textit{et al.} \cite{Torikachvili08} as well as by Kumar \textit{et al.} \cite{Kumar08} agree with our data well, but neither of these studies entered into the zero resistivity dome. Higher pressure transport measurements made by Kotegawa \textit{et al.} \cite{Kotegawa09} suppress the structural / antiferromagnetic phase transition in a manner similar to our results and reach pressures high enough (P $\thicksim$ 43 kbar) to enter into the zero resistivity dome. Unfortunately these measurements do not go to high enough pressures to suppress the zero resistivity dome. Susceptibility measurements by Alireza \textit{et al.} \cite{Alireza08} detected diamagnetism over a similar pressure range as our zero resistivity dome with a local maximum in diamagnetic signal close to the pressure where we detect a maximum in T$_{c}$ for the dome. On the other hand, the T$_{c}$ values found by Alireza \textit{et al.} \cite{Alireza08} suddenly become finite at T $\thicksim$ 27~K on the low pressure side of this region and monotonically drop to about 20~K on the high pressure side (in a very non-linear fashion). The one exception to the general agreement about the pressure dependence of the structural / antiferromagnetic phase line is transport work by Igawa \textit{et al.} \cite{Igawa08} that shows a dramatically slower suppression of the resistive feature. This work was done on a polycrystalline sample, and for highest pressures used NaCl as a pressure medium. Given the known sensitivity of these materials to strain, it is not surprising that liquid media and single crystals are preferable.

In the case of \BFAf , the literature is more sparse. There are two transport studies: Fukazawa \textit{et al.} \cite{Fukazawa08} on polycrystalline samples in a liquid medium and Mani \textit{et al.} \cite{Mani09condmat} on poly- and single-crystalline samples in a solid medium (steatite), both scenarios are prone to strain. Mani \textit{et al.} \cite{Mani09condmat} cannot detect a structural / antiferromagnetic transition for pressures greater than P $\thicksim$ 15~kbar and a qualitative comparison between pressures is not possible, due to probable cracks / irreversible defects to the sample as their strong increase of the ambient temperature resistivity (higher than a factor 5 between 0 and 72~kbar) proved. Fukazawa \textit{et al.} \cite{Fukazawa08} find a suppression of this phase transition that is much slower than our results. Fukazawa \textit{et al.} \cite{Fukazawa08} result is similar in its deviation from single crystal results as the polycrystalline work by Igawa \textit{et al.} \cite{Igawa08} on \SFA discussed above. Susceptibility measurements by Alireza \textit{et al.} \cite{Alireza08} detected diamagnetism but at pressures essentially shifted by 10~kbar lower than our zero resistivity dome. Since Alireza \textit{et al.} \cite{Alireza08} did not measure the pressure dependence of the structural / antiferromagnetic phase transition it is impossible to determine how well their superconducting region correlates with the higher temperature phase line.

For both the \SFA and the \BFA systems studies of the transport properties of single crystals under high, but as hydrostatic as possible, pressure are needed to establish the relation between the structural / antiferromagnetic phase line and the superconducting (or zero resistivity) dome. In our studies of these systems up to P $\thicksim$ 80~kbar we have been able to achieve this goal.

The relationship between the low temperature kink, seen in our data at all pressures higher than ambient, and zero resistivity (at least complete filamentary superconductivity and possible bulk superconductivity) is clearly described by our data and summarized in the two T(P) phase diagrams shown in Figs. \ref{DP} and \ref{DPBa}.  For both compounds it is worth noting that the temperature associated with a zero resistance state is relatively pressure sensitive, rising toward the kink temperature and then dropping away from it as pressure is increased. These data support the idea that the kink-like feature can be associated with some form of strain induced superconductivity in a very small fraction (below the percolation limit) of the sample with a distribution of T$_{c}$ values ranging as high as the maximum T$_{c}$ for the material. Once induced this hypothetical strain field is relatively pressure insensitive. This is consistent with recent work by Saha \textit{et al.} \cite{Saha08condmat} for the case of \SFA where high temperature annealing is necessary to remove the strain induced kink and superconductivity.

\begin{figure}
\begin{center}
\includegraphics[angle=0,width=120mm]{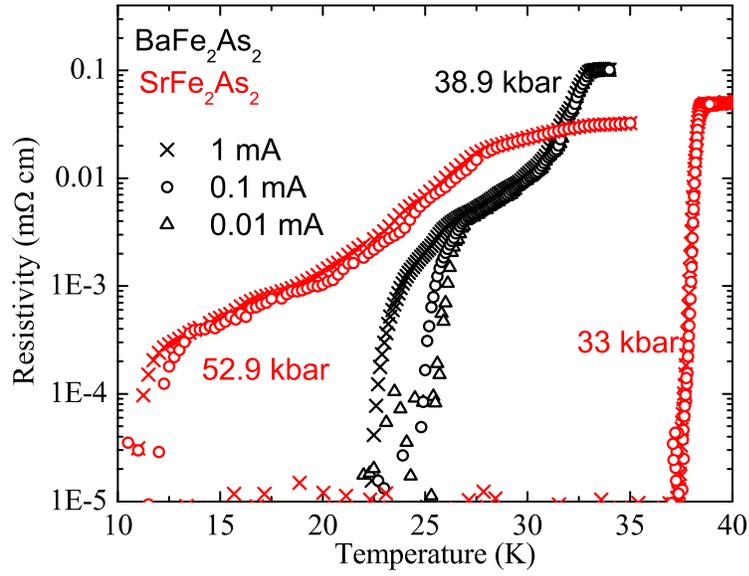}
\end{center}
\caption{(Color online) Superconducting transition measured in resistivity shown for different currents: 0.01, 0.1 and 1 mA for three pressures. \SFA at 33 and 52.9~kbar and \BFA at 38.9 kbar.}
\label{current}
\end{figure}

The degree of superconductivity associated with the zero resistivity dome can be probed (at least a little) by the measurement of current dependent resistivity. Figure \ref{current} presents data for \BFA with P $\thicksim$ 39~kbar (near the onset of the zero resistivity dome), and for \SFA for P $\thicksim$ 33~kbar (near the local maximum of the zero resistivity dome), and P $\thicksim$ 53~kbar (on the high pressure side of the local maximum of the zero resistivity dome). Whereas there is no significant current dependence of the resistivity for pressures near the optimal pressure for superconductivity, there is a clear current dependence on both the low and high pressure sides. This is consistent superconductivity of a more filamentary nature existing at the low and high pressure edges of this dome. Unfortunately the lack of a significant current dependence (over this limited current range) does not prove true, bulk superconductivity in the sample, even at the optimal pressure, but the data does allow for the comparative statement that the superconductivity is less filamentary near the center of the zero resistivity dome.

The location of the zero-resistivity dome in the \BFA and \SFA T(P) phase diagrams is noteworthy as well. For each of these compounds we find that the maximum in T$_{c}$ is found near the pressure where the extrapolation of the structural / antiferromagnetic phase line reaches zero. (This key observation is possible because we were able to detect both the upper and lower phase transitions during the same measurement.) Extrapolations of the structural / antiferromagnetic phase line gives a critical pressure of P $\thicksim$ 35~kbar for \SFA and P $\thicksim$ 55~kbar for \BFA (both of which match the maximal T$_{c}$ regions nicely). Although, as discussed above, the range of bulk superconductivity is not well know, the central region of the zero-resistivity dome is the most likely pressure range to find bulk superconductivity.

\begin{figure}
\begin{center}
\includegraphics[angle=0,width=120mm]{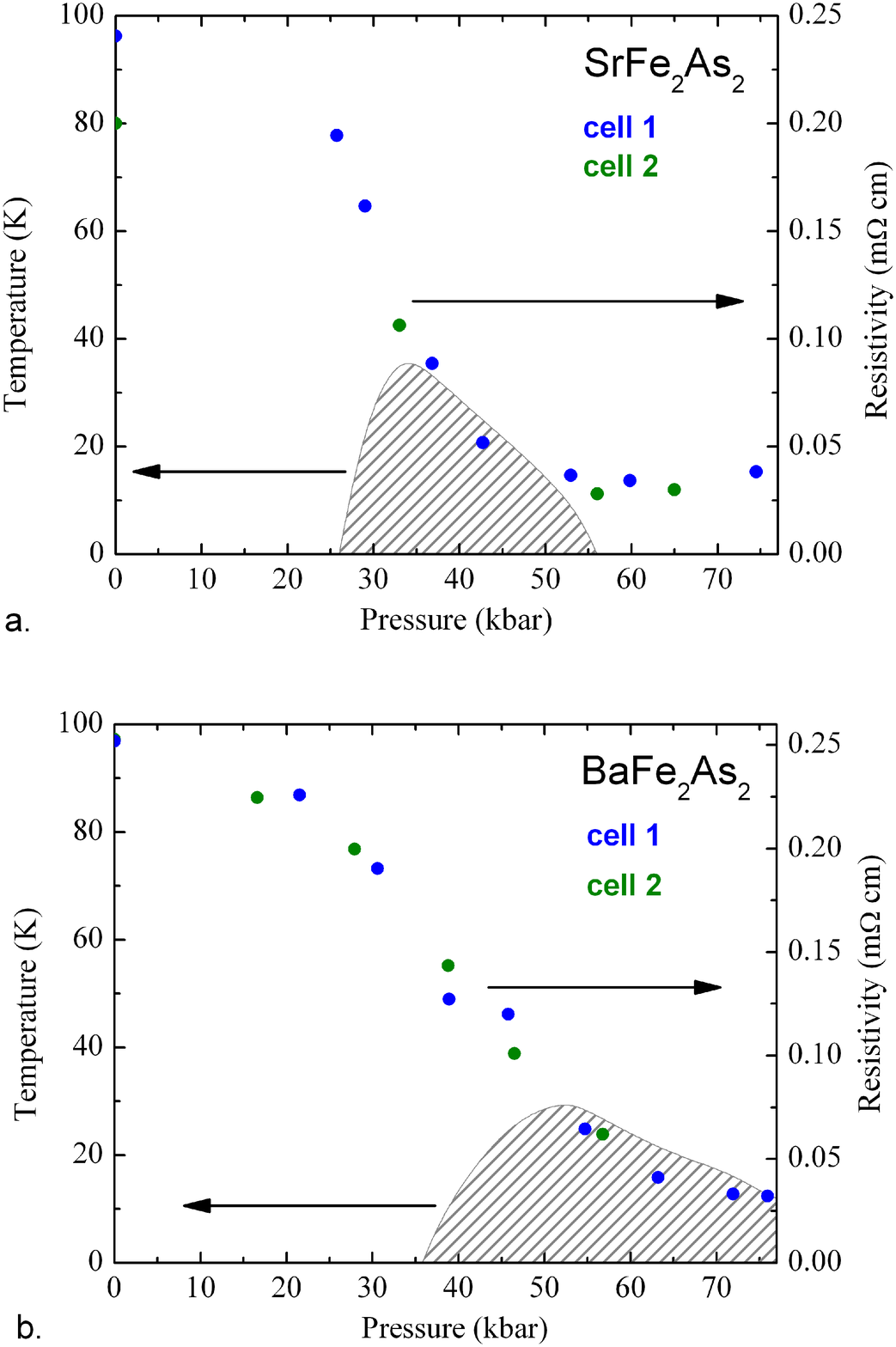}
\end{center}
\caption{(Color online) Pressure dependence of the resistivity at 40~K (right axis) (a) for \SFA and (b) for \BFAf . For reference, the zero-resistivity dome (hatched area) from the T(P) phase diagram is also shown (left axis).}
\label{DPwithres}
\end{figure}

The location of the superconducting dome can also be related to the changing, low temperature, normal state resistivity. Figure \ref{DPwithres} presents the low temperature resistivity (just above the maximum superconducting or kink temperature) as a function of pressure for \SFA (a) and \BFA (b).  For reference Fig. \ref{DPwithres} also shows the location of the zero resistivity dome. For \SFA (with its lower characteristic pressure scale) the R$_{40\text{K}}$(P) data manifest the sharpest drop right at the pressure associated with the maximum in the zero-resistivity dome. This is also the pressure range that the extrapolation of the structural / antiferromagnetic phase line would cross 40~K. For \BFA (with a higher characteristic pressure) a similar correlation between the change in the low temperature resistivity and the zero resistivity dome can be observed, but at higher pressures and over a wider pressure range.

The location of the zero resistivity dome around the critical pressure for the structural / antiferromagnetic phase transition raises the question of possible quantum criticality.  Unfortunately we only have resistivity data and, given the relatively high H$_{c2}$(T) values for these materials (as shown in Figs. \ref{field} and \ref{Bafield}), we only have these data for relatively high temperature (T $\geqslant$ 40~K near the critical pressure). This being said, a closer examination of the temperature dependent resistivity data presented in Figs. \ref{cell} and \ref{cellBa} reveals a more linear-like temperature dependence just above T$_{c}$ for pressures close to the optimal pressure and more super-linear temperature dependences for pressures both below and above the optimal pressure. Measurements in a diamond anvil cell, with He as a pressure media and exceptionally high magnetic fields (H $\lesssim$ 60~T) together with the normal state magneto-resistive corrections will be needed to make a more quantitative statement.

As the pressure is increased beyond the critical pressure the $\rho$(T) curves start to fall on a universal, low temperature manifold, with each subsequent pressure defining this manifold to a higher temperature. This behavior is again more easily observed in \SFA given its lower critical pressure, although is can be observed starting to set in for the highest pressure \BFA data sets as well. This behavior is reminiscent to what was observed in \CFA as increasing pressure stabilized the collapsed tetragonal phase at higher and higher temperatures. For \SFA and \BFA there is no evidence for a phase transition to a collapsed tetragonal phase, but there is evidence for a pressure stabilized high temperature state that has greatly reduced resistivity. For each of the \AFA (AE = Ca, Sr, and Ba) this low resistivity state does not support superconductivity. This may well be related to the more general observation that can be made about superconductivity in the doped FeAs-based compounds:  superconductivity occurs when the fluctuations or correlations associated with the high temperature tetragonal state are brought to "low enough" temperature. If these fluctuations or correlations are fully suppressed (i.e. the resistivity is fully reduced to that of a non magnetic, non-correlated metal) then superconductivity is no longer supported.

\section{Conclusions}
By measuring several samples of \BFA and \SFA in a liquid medium, self clamping, Bridgman cell up to pressures approaching 80~kbar we have been able to determine the complete pressure - temperature phase diagrams for these two parent compounds of the \AFA superconductors. Both of these T(P) diagrams consist of three basic features, (i) a structural / antiferromagnetic phase transition that is suppressed by increasing pressure, (ii) a zero resistivity dome that is relatively pressure sensitive and also appears to represent less filamentary superconductivity near its central region, and (iii) a kink-like feature that is relatively pressure insensitive that is thought to be associated with small parts of the sample manifesting a spread of T$_{c}$ values, probably originating from strains / defects rather than from the hydrostatic pressure. We have found that the zero resistivity dome is centered around the critical pressure for the structural / antiferromagnetic phase transition (P $\thicksim$ 35 kbar for \SFA and P $\thicksim$ 55 kbar for \BFAf). We have determined this critical pressure both via the extrapolation of the structural / antiferromagnetic phase line down to T = 0~K and via the change in the low temperature (40~K), normal state resistivity associated with transition temperature passing through T = 40~K.\\

These data imply that the superconductivity found in this system may be linked to a quantum critical point associated with the suppression of the structural / antiferromagnetic phase transition. Although the high temperature, and high H$_{c2}$(T) curves, associated with the superconductivity make quantitative analysis of the resistivity difficult, there does appear to be a trend toward more linear-like temperature dependence of the resistivity in the region of this critical pressure and more super linear temperature dependences for both lower and higher pressures.
\\

More quantitatively we can link superconductivity in \SFA and \BFA to bringing the fluctuations / correlations associated with the low pressure tetragonal state to low enough temperatures. The zero resistivity dome exists in the region of phase space where the structural / antiferromagnetic phase is suppressed to low enough temperatures and the fluctuations / correlations associated with the tetragonal phase (as measured by the resistivity) are not completely suppressed. To this end \SFA and \BFA under pressure appear to manifest the same basic physics as doped \BFA \cite{MulipleTmDoping}, but with a different tuning parameter.

\begin{acknowledgments}
Work at the Ames Laboratory was supported by the Department of Energy, Basic Energy Sciences under Contract No. DE-AC02-07CH11358. We would also like to acknowledge N. H. Sung for his assistance in pressure cells preparation, and Daniel Braithwaite and CEA Grenoble for helping establish the liquid media, self clamping Bridgman cell in Ames Laboratory.
\end{acknowledgments}

\clearpage
\clearpage
\clearpage
\clearpage
\clearpage
\clearpage
\clearpage
\clearpage
\clearpage
\begin{figure}
\begin{center}
\includegraphics[angle=0,width=120mm]{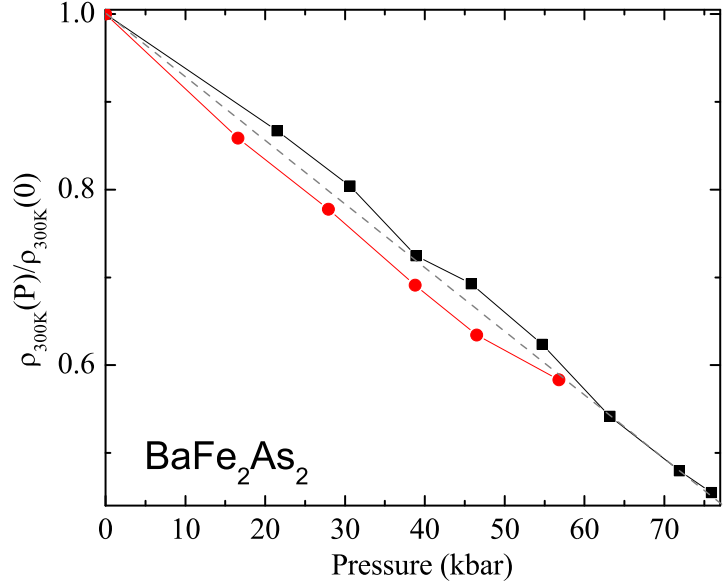}
\end{center}
\caption{(Color online) Relative resistivity decrease at 300~K $\rho_{300\text{K}}(P)/\rho_{300\text{K}}(0)$ versus pressure from cell 1 (black squares) and cell 2 (red circles). The dashed line is a guide for the eye. }
\label{R300KBa}
\end{figure}
\clearpage
\begin{figure}
\begin{center}
\includegraphics[angle=0,width=120mm]{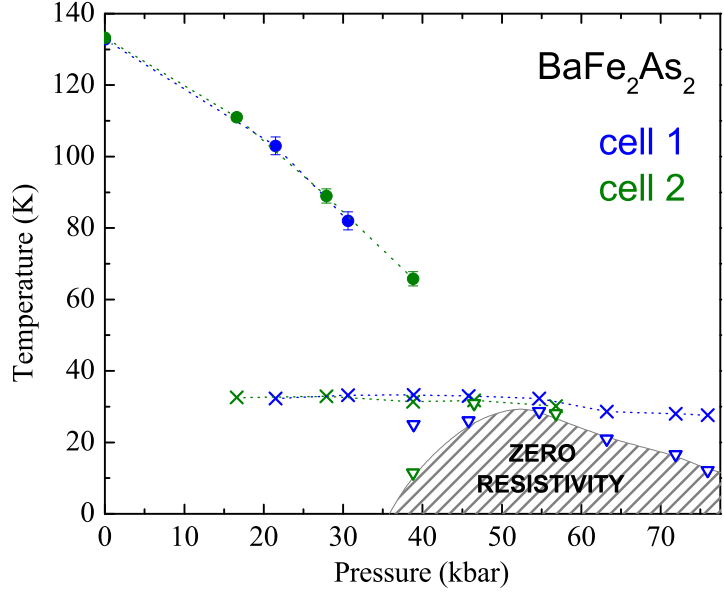}
\end{center}
\caption{(Color online) Phase diagram T(P) of \BFA deduced from resistivity measurements in a modified Bridgman pressure cell. Green and blue colors refer to two different cells, respectively called as cell 1 and cell 2. Circles correspond to the structural transition deduced from the local maximum of the resistivity derivative. Vertical error bars indicates the shift between cooling and warming, due to a not perfect thermalization. Crosses correspond to the onset of the superconducting transition. Triangles represent offset temperature of the full superconducting transition. The hatched area estimates the true zero-resistance superconducting region.}
\label{DPBa}
\end{figure}
\clearpage
\clearpage
\clearpage

\end{document}